\documentclass[aps,prl,epsfig,showpacs,superscriptaddress,footnote,twocolumn]{revtex4}
\makeatletter
\usepackage{calc}
\usepackage{times}
\usepackage{multirow}
\def\be{\begin{equation}}
\def\ee{\end{equation}}
\def\bea{\begin{eqnarray}}
\def\eea{\end{eqnarray}}
\def\bma{\begin{mathletters}}
\def\ema{\end{mathletters}}
\def\bi{\begin{itemize}}
\def\ei{\end{itemize}}

\def\C{\hbox{$\mit I$\kern-.7em$\mit C$}}

\newcommand{\singlespacing}{\let\CS=\@currsize\renewcommand{\baselinestretch}
{1.0}\tiny\CS}
\newcommand{\doublespacing}{\let\CS=\@currsize\renewcommand{\baselinestretch}
{1.5}\tiny\CS}

\begin{document}
\title{Bound on Hardy's non-locality from
the principle of Information Causality}

\author{Ali Ahanj}
\email{ahanj@ipm.ir} \affiliation{Khayyam Institute of Higher
Education, Mashhad, Iran}\affiliation{ School of Physics,
Institute for Research in fundamental Sciences, P.O.Box
19395-5531, Tehran, Iran}

\author{Samir Kunkri}
\email{skunkri@yahoo.com} \affiliation{Mahadevananda
Mahavidyalaya, Monirampore, Barrackpore, North 24 Parganas,
700120, India }

\author{Ashutosh Rai}
\email{arai@bose.res.in} \affiliation{S.N.Bose National Center for
Basic Sciences,Block JD, Sector III, Salt Lake, Kolkata-700098,
India}

\author{Ramij Rahaman}
\email{ramij.rahaman@ii.uib.no} \affiliation{Selmer Center,
Department of Informatics, University of Bergen, Bergen, Norway}

\author{Pramod S. Joag}
\email{pramod@physics.unipune.ernet.in} \affiliation{Department of
Physics,University of Pune, Ganeshkhind, Pune-411007, India}

\begin{abstract}
Recently, the principle of non-violation of Information Causality
[Nature 461, 1101 (2009)], has been proposed as one of the
foundational properties of nature. We explore the Hardy's
non-locality theorem for two qubit systems, in the context of
 generalized probability theory, restricted by the principle of
 non-violation of Information Causality. Applying, a sufficient
 condition for Information causality violation, we derive an upper
 bound on the maximum success probability of Hardy's nonlocality
 argument. We find that the bound achieved here is higher than
 that allowed by quantum mechanics, but still much less than what
 the no-signalling condition permits.
 We also study the Cabello type non-locality argument (a generalization of Hardy's
argument) in this context.
\end{abstract}

\pacs{03.65.Nk, 03.65.Yz}

\maketitle
\section*{Introduction}
Understanding quantum non-locality is a fundamental problem facing
science \cite{nonlocality}. Quantum non-locality is profoundly
manifested via the correlations between the measured values of
physical quantities pertaining to different (spatially separated)
parts of the quantum system, which cannot be generated locally.
That is, quantum correlations violate Bell's inequality
\cite{bell}, which any local hidden variable theory has to
satisfy. Despite this, quantum non-locality cannot be used for
signalling between distant quantum systems \cite{ghirardi}.\\

In order to quantify and explain the effects of quantum
non-locality, we have to relate it to some fundamental
principle(s). These principle(s) must ensue from violation of Bell
type inequalities (a basic manifestation of quantum non-locality),
and the no-signalling principle. One such approach, initiated by
Popescu and Rohrlich \cite{PR}, is to find the class of models
following exclusively from non-locality and no-signalling
requirements, and then ask which of them coincides with quantum
mechanics. They showed that it is possible to write down sets of
``superquantum`` correlations that are more nonlocal than that
allowed by quantum mechanics, yet satisfying nonsignalling. They
invented a hypothetical device now known as PR box, which
generates a set of correlations that return a value of $4$ for the
Clauser-Horne-Shimony-Holt (CHSH) expression \cite{CHSH}, the
maximum algebraic value possible, yet are nonsignaling. The
maximum quantum value is given by Tsirelson's theorem
\cite{tsirelson} as $2\sqrt{2}$ . Thus only the no-signalling
requirement does not give a non-local theory which uniquely
determines quantum mechanics. Several important works
\cite{a,b,c,d,e,f,g} thereafter, show that numerous properties
which were usually thought to be typically quantum, are also
shared by general nonsignalling models. Although, W. van Dam
\cite{h} showed that the strongest nonsignalling  correlations
(viz. PR boxes) make communication complexity trivial, which seems
highly implausible in Nature. An intriguing question to ask is,
what are characterizing features of quantum mechanics
(Nature), which makes it unique.\\

Very recently, Pawlowaski et al \cite{IC} give a bold new proposal
for understanding the quantum mechanical bound on nonlocal
correlations. They showed that if non-violation of Information
Causality (IC) is accepted as a foundational principle of nature,
then it follows that, a class of nonlocal correlations which are
postquantum violate this principle, while Quantum and Classical
correlations respect it. The principle states that communication
of $m$ classical bits causes information gain of at most $m$ bits,
this is a generalization of the no-signalling principle, the case
$m=0$ corresponds to no-signalling. On applying IC principle to
non-local correlations, Tsirelson's bound \cite{tsirelson}
naturally emerges, all correlations exceeding Tsirelson's bound
violate the principle of information causality. However, there are
also nonquantum correlations which are bounded by $2\sqrt{2}$
(Tsirelson's bound), so it remains interesting to see whether
various quantum correlations follows from IC condition or not.
Allcock et al \cite{allcock} show that a part of the boundary
between quantum and nonquantum correlations can indeed be
recovered from the principle of Information Causality. Studying
quantum nonlocality from different perspectives (like non-locality
without inequalities), in the context of IC may lead to a deeper
understanding of the issues involved.\\

Hardy \cite{hardy92, hardy93} gave a non-locality theorem (in
contrast to Bell's inequality) which provides a manifestation of
quantum non-locality, without using statistical inequalities
involving expectation values. Later, Cabello \cite{cabello02} has
introduced a logical structure to prove Bell's theorem without
inequality for three particles GHZ and W states and the argument
is also applicable for two qubits \cite{kunkri,liang}. Hardy's
logical structure is a special case of Cabello's structure. The
maximum probability of success of Hardy's non-locality argument
for two qubit systems is $0.09$ whereas in case of Cabello's
argument it is $0.1078$ \cite {kunkri}. Maximum success
probability of Hardy's nonlocality argument for the class of
generalized no signalling theories in two qubit systems is $0.5$
\cite{GNLT,cereceda}.

Barrett et al \cite{barrett} with a aim to shed light on the range
of quantum possibilities by placing them in a wider context
(generalized probability theory), investigated the set of
correlations that are constrained only by the no-signaling
principle. They found that these correlations form a polytope,
which contains the quantum correlations as a (proper) subset.
Their work sets an elegant mathematical framework for
understanding the general structure of these nonlocal correlations. \\

In the present paper we study Hardy's non-locality arguments for
two qubit systems in the context of non-violation of information
causality. We adopt the characterization of no signalling boxes
and notations used in Barrett et al \cite{barrett}. We have found
that unlike CHSH inequality, Hardy's bound on the success of
non-local probabilities cannot be recovered from the condition
\cite{IC} which follows from the principle of non-violation of information causality.\\

\section*{Hardy's/Cabello's argument for two qubits}
Consider two spin-1/2 particles $1$ and $2$ with spin observable
$A$, $A^{\prime}$ on particle $1$ and $B$, $B^{\prime}$ on
particle $2$. These observable gives the eigenvalues $\pm{1}$. Now
consider the following joint probabilities:
\begin{eqnarray}
P(A = +1, B = +1) &=& q_1\\
P(A^{\prime} = -1, B = -1) &=& 0\\
P(A = -1, B^{\prime} = -1) &=& 0\\
P(A^{\prime} = -1, B^{\prime} = -1) &=& q_4
\end{eqnarray}
Here equation (1) tells that, if A is measured on particle 1 and B
is measured on particle 2, then the probability that both get
value +1 is $q_1$, remaining equations can also be interpreted in
a similar fashion. These equations form the basis of Cabello's
nonlocality argument. It can easily be seen that these equations
contradict local-realism if $q_1 < q_4$. To show this, let us
consider those hidden variable states $\lambda$ for which
$A^{\prime} = -1$ and $B^{\prime} = -1$. For these states,
equations $(2)$ and $(3)$ tell that the values of $A$ and $B$ must
be equal to $+1$. Thus according to local realism $P(A = +1, B =
+1)$ should be at least equal to $q_4$. This contradicts equation
$(1)$ as $q_1 < q_4$. It should be noted here that $q_1=0$ reduces
this argument to that of Hardy's. So by Cabello's argument, we
specifically mean that the above argument runs, even with nonzero
$q_1$.
\section*{Hardy's correlations from no-signalling polytope}
Given a set of observables $X, Y \in {\{0~~ 1\}}$ and outcomes $a,
b \in {\{0~~ 1\}}$ joint probabilities $p_{ab|XY}$ thus form an
entire correlation table with $2^4$ entries, which can be regarded
as a point of $2^4$- dimensional vector space. The positivity,
normalization and non-signalling constraints lead the entire
correlation table to a convex subset in the form of a polytope
which is known as no-signalling polytope ${\cal P}$, which is
eight dimensional \cite{barrett}. There are $24$ vertices of the
polytope ${\cal P}$, $16$ of which represent local correlations (
called ``local vertices") and $8$ represent nonlocal correlations.
The local vertices can be expressed as
\begin{equation}
\label{local} p_{ab|XY}^{\alpha \beta \gamma \delta} =
\left\{\begin{array}{ccccc}
                   1, & {\rm if} & a &=& {\alpha}X \oplus {\beta},\\
                     &           & b &=& {\gamma}Y \oplus {\delta};\\
                   0, & {\rm otherwise} & & &
                   \end{array}
             \right.
\end{equation}
where $\alpha, \beta, \gamma, \delta \in \{ 0, 1\}$ and $\oplus$ denotes addition modulo $2.$\\
 The eight nonlocal vertices have the form:

\begin{equation}
\label{nonlocal} p_{ab|XY}^{\alpha \beta \gamma } =
\left\{\begin{array}{ccccc}
                   \frac{1}{2}, & {\rm if} & a \oplus b &=& XY \oplus {\alpha}X \oplus {\beta}Y \oplus \gamma,\\
                   0, & {\rm otherwise} & & &
                   \end{array}
             \right.
\end{equation}
where $\alpha, \beta, \gamma \in \{ 0, 1\}$.\\
Now we use the correspondence $(X=0)\leftrightarrow
A,(X=1)\leftrightarrow A^{\prime},$ $(Y=0)\leftrightarrow
B,(Y=1)\leftrightarrow B^{\prime}$ and $a,b=0(1)\leftrightarrow
+1(-1).$ Then it is straight-forward to check that five of the
$16$ local vertices and one of the $8$ nonlocal vertices satisfy
Hardy's equations $(1)$-$(4)$ (when $q_1 = 0$), namely those given
by $p_{ab|XY}^{0001}$, $p_{ab|XY}^{0011}$, $p_{ab|XY}^{0100}$,
$p_{ab|XY}^{1100}$, $p_{ab|XY}^{1111}$ and $p_{ab|XY}^{001}$.
 The other vertices can be covered by another set of Hardy's equations.
 Then the joint probabilities satisfying Hardy's conditions can be written as a convex
 combination of the above $6$ vertices (five local vortices and one nonlocal vertex). Then
\begin{eqnarray}
\label{hardy} p_{ab|XY}^{{\cal H}} &=& c_1 p_{ab|XY}^{0001} + c_2
p_{ab|XY}^{0011} + c_3 p_{ab|XY}^{0100} \nonumber \\
&&+ c_4 p_{ab|XY}^{1100}+ c_5 p_{ab|XY}^{1111} + c_6
p_{ab|XY}^{001}
\end{eqnarray}
where $\sum_{j = 1}^{6} {c}_i = 1.$\\ Now if we consider $q_1 \ne
0$ (but $q_1 < q_4$), then the equations $(1)-(4)$ is known as
Cabello's nonlocality conditions, which can be written as a convex
combination of the above $6$ vertices which satisfies Hardy's
conditions along with another four local vertices
$p_{ab|XY}^{0000}$, $p_{ab|XY}^{0010}$, $p_{ab|XY}^{1000}$,
$p_{ab|XY}^{1010}$ and one nonlocal vertex $p_{ab|XY}^{110}$. So
we get,
\begin{eqnarray}
\label{cabello} p_{ab|XY}^{{\cal C}} &=& p_{ab|XY}^{{\cal H}} + c_7
p_{ab|XY}^{0000} + c_8 p_{ab|XY}^{0010} + c_9 p_{ab|XY}^{1000}
\nonumber \\
&&+ c_{10} p_{ab|XY}^{1010} + c_{11} p_{ab|XY}^{110}
\end{eqnarray}
where the expression $p_{ab|XY}^{{\cal H}}$ is given in equation
(\ref{hardy}) and coefficients $c_i$'s satisfy the condition $\sum_{j = 1}^{11} {c}_i = 1.$\\
One can check from equation (\ref{hardy}) that the success
probability for Hardy's argument is given by $p_{11|11}^{{\cal H}}
= \frac {1}{2}c_6$. From here, one can obviously see that under
the no-signaling constraint, the maximum success probability of
Hardy's argument {\it i.e.} $(p_{11|11}^{{\cal H}})_{max}=
\frac{1}{2}$ is achieved for $c_{6}=1$ and
$c_{1}=c_{2}=c_{3}=c_{4}=c_{5}=0 $. Similarly the success
probability for Cabello's argument follows from equation
(\ref{cabello}) and can be written as, $p_{11|11}^{{\cal
C}}-p_{00|00}^{{\cal C}}=(\frac{1}{2}c_{6}+ c_{10})-C$, where
$C=c_{7}+c_{8}+c_{9}+c_{10}+\frac{1}{2}c_{11}$, and, here too we
obtain that $(p_{11|11}^{{\cal C}}-p_{00|00}^{{\cal C}})_{max}=
\frac{1}{2}$ for $c_{6}=1$ and rest of the $c_{i}$'s$=0$. This
maximum success probability of Hardy's/Cabello's argument,
restricted by the no-signalling condition, has also been derived
in \cite{GNLT,cereceda}. One should note that the probability set
for which this maximum is achieved coincides with PR correlation
for both the cases.
 In the following sections we will derive an upper bound on
the maximum value of these success probabilities from the
principle of non-violation of information causality.

\section*{Information Causality}
Let us now  briefly review the principle of information causality
(IC). Alice and Bob, who are separated in space, have access to
non-signalling resources such as shared randomness, entanglement
or (in principle) PR boxes. Alice receives $N$ random bits $\vec
{a} = (a_1, a_2,....., a_N)$ while Bob receives a random variable
$b \in \{1, 2,, ..., N\}$. Alice then sends $m$
 classical bits to Bob, who must output a single bit $\beta$ with the aim of
 guessing the value of Alice's b-th bit $a_b$. Their degree of success at this
 task is measured by $$ I \equiv \sum_{K = 1}^{N} {I (a_K : \beta |b = K)},$$
 where $I (a_K : \beta |b = K)$ is Shannon mutual information between $a_K$ and $\beta$.
  The principle of information causality states that physically allowed theories must have

\begin{equation}
\label{causality}
I \le m.
\end{equation}

It was proved in \cite{IC} that both classical and quantum
correlations satisfy this condition.  A condition under which
IC is violated was derived in \cite{IC}, based on a construction by Van Dam and Wolf and
 Wullschleger- for a specific realization of the Alice-Bob channel.
 It goes as follows. Define $P_1$ and $P_2$:
\begin{eqnarray}
\label{pipii}
P_1&=& \frac{1}{2}\left[p_{(a=b|00)} + p_{(a=b|10)}\right]\nonumber\\
   &=& \frac{1}{2}\left[p_{00|00} + p_{11|00} + p_{00|10} + p_{11|10}\right]\nonumber\\
P_2&=& \frac{1}{2}\left[p_{(a=b|01)} + p_{(a\neq b|11)}\right]\nonumber\\
   &=& \frac{1}{2}\left[p_{00|01} + p_{11|01} + p_{01|11} + p_{10|11}\right]
\end{eqnarray}
Then, the IC condition (\ref{causality}) is violated for all boxes for which
\begin{equation}
\label{ic} E^{2}_1 + E^{2}_2 >1,
\end{equation}
where $E_j= 2P_j-1$ ($j=1,2$). Here it is important to note that
the condition (\ref{ic}) is only a sufficient condition (based on
the protocol give in \cite{IC}) for violating the IC principle.
\section*{Hardy's nonlocality and Information Causality}
In this section we derive an upper bound on the maximum
probability of success of Hardy's non-locality argument for a two
qubit system in the context of non-violation of information
causality. Let Alice and Bob share non-signalling nonlocal
correlation satisfying Hardy's condition {\it i.e.} the joint
probability $P_{ab|XY}^{\cal H}$ given in equation (\ref{hardy}).
Then for this nonlocal correlation we have
\begin{eqnarray}
\label{p1}
P_1 &=& \frac{1}{2}(c_5 + c_4),\nonumber\\
P_2 &=& \frac{1}{2}(c_1 + c_2 + c_3)
\end{eqnarray}
To satisfy the IC condition equation (\ref{p1}) has to satisfy the condition
$$ E^{2}_1 + E^{2}_2 \leq 1 $$ {\it i.e} $$(c_5 + c_4 -1)^2 + (c_1 + c_2 + c_3 -1)^2 \leq 1,$$ which implies
\begin{eqnarray}
\label{pc}
c_6^2 + 2 (c_4 + c_5) c_6 + 2 (c_4 + c_5)(c_4 + c_5 - 1) \leq 0
\end{eqnarray}
The above equation gives the maximum value of $c_6 = \sqrt{2}-1$.
Then an upper bound on the maximum probability of success of
Hardy's non-locality is given by $P_{11|11}^{\cal H} = \frac
{1}{2}c_6 \leq \frac{1}{2}(\sqrt{2}-1)=0.20717.$

\section*{Cabello's nonlocality and Information Causality}
Now we try to find an upper bound on the maximum probability of
success in Cabello's case in the context of non-violation of
information causality. Let Alice and Bob share non-signalling
nonlocal correlation satisfying Cabello's condition i.e joint
probability $p_{ab|XY}^{{\cal C}}$ given in equation
(\ref{cabello}). Then for this nonlocal correlation we have:
\begin{eqnarray}
\label{p1c} P_1 &=& \frac{1}{2}[C + c_5 + \frac{c_{11}}{2}  + c_4
+c_7 + c_8]\nonumber\\
P_2 &=& \frac{1}{2}[1 + c_9 - (c_4 + c_5 + c_6 + c_{10})]
\end{eqnarray}
where $ C = c_7 +c_8 + c_9 + c_{10} + \frac{1}{2}c_{11}$. Then
\begin{eqnarray}
\label{e1c} E_1 &=& c_7 + c_8 - c_1 - c_2 - c_3 - c_6 \nonumber\\
E_2 &=& c_9 - c_4 - c_5 - c_6 - c_{10}
\end{eqnarray}
To satisfy the IC condition equation (\ref{e1c}) has to satisfy
the condition $$ E^{2}_1 + E^{2}_2 \leq  1. $$ One can easily
check that  $$ E_1 + E_2 = -(1 + 2x)$$ where $x = (c_{10} +
\frac{1}{2}c_{6}) - C.$ It follows that $$ E_1^2+E_2^2 = 4x^2 +
4(1+E_2)x + 2(1+E_2)E_2 + 1, $$ so in order to satisfy the IC
conditon,
$$x^2+(1+E_2)x+\frac{1}{2}E_2(1+E_2)\leq 0$$
writing $E_2$ in terms of $P_2$ we obtain,
$$ x^2+2P_2x+P_2(2P_2-1)\leq 0.$$ Since we are to find $x_{max}$
it is sufficient to consider only the equality. Then,
$$ x=-P_2+\sqrt{P_2(1-P_2)};$$
 $$0\leq P_2 \leq \frac{1}{2}.$$
The maximum value of $x$ we obtained from here is,
$$x_{max}=\frac{1}{2}(\sqrt{2}-1)=0.20717.$$
This value is same as in the Hardy's case. We conclude that on
applying the IC condition, maximum probability of success of the
Cabello's argument is same as that of the Hardy's argument, both
achieving the same numerical value $0.20717$.
\section*{Conclusion}
The maximum probability of success of the Hardy's and Cabello's
non-locality (for the two qubits system) in Quantum mechanics is
$0.09$ and $0.1078$ respectively \cite{kunkri}. Interestingly, for
generalized nonlocal no-signalling theories we find that this
bound is $0.5$ in both the cases and the probability set for which
this is achieved coincides with the PR correlation. We showed that
on applying the principle of Information causality this bound
decreases from $0.5$ to $0.20717$ in both the cases, but could not
reach their respective Quantum mechanical bounds. Interestingly,
in quantum mechanics the maximum probability of success for the
Cabello's case is not same as the Hardy's case \cite{kunkri}.
Since the condition given by equation (\ref{ic}) of the present
paper is a sufficient condition derived in \cite{IC} for the
violating IC, the probability derived here are therefore,
strictly, an upper bound on the maximum probabilities allowed in
an IC-respecting no-signaling theory. Restricting the
no-signalling probability set by the full power of IC principle
may reduce the probability to the quantum limit. However, it is
curious that the same sufficient
condition for violating the IC, gives the quantum bound for the CHSH expression \cite{IC}.\\ \\ \\ \\

\textbf{ACKNOLEDGMENTS}

It is a pleasure to thank Guruprasad Kar for many stimulating
discussions and encouragement. SK thanks Sibasish Ghosh for
several discussions during his visit at IMSc (Chennai). AR
acknowledge support from DST project SR/S2/PU-16/2007. RR
acknowledge support from Norwegian Research Council. This work was
partially carried out when AA and PSJ visited ISI (Kolkata). They
thank ISI (Kolkata) for the invitation and hospitality and for its
conducive research environment.

\end{document}